\documentclass{aastex}

\usepackage{spr-astr-addons}
\usepackage{url}

\RequirePackage{color}

\usepackage{multirow}
\def\rot{\rotatebox}

\begin{document}

\title{Periodicity in  the continua and broad  line curves of a quasar  E1821+643}
\slugcomment{Manuscript}
\shorttitle{Periodicity of E1821+643 light curves}
\shortauthors{Kova{\v c}evi{\'c} et al.}

\author{A.  Kova{\v c}evi{\'c}\altaffilmark{1}} 
\affil{Department of Astronomy, Faculty of Mathematics, University of Belgrade, Studentski trg 16, 11000 Belgrade, Serbia, \\ andjelka@matf.bg.ac.rs, tel 381648650032, fax 381112630151}
\and \author{L. {\v C} Popovi{\'c}\altaffilmark{1,2}}
\affil{Astronomical Observatory Belgrade, Volgina 7, 11060 Belgrade, Serbia, lpopovic@aob.rs,\\ Department of Astronomy, Faculty of Mathematics, University of Belgrade, Serbia}
\and
\author{A. I. Shapovalova \altaffilmark{3}}
\affil{Special Astrophysical Observatory of the Russian AS, Nizhnij Arkhyz, 
Karachaevo - Cherkesia 369167, Russia,  ashap@sao.ru}
\and
\author{D.  Ili{\'c} \altaffilmark{1}}
\affil{Department of astronomy, Faculty of mathematics, University of Belgrade, Studentski trg 16, 11000 Belgrade, Serbia, dilic@matf.bg.ac.rs  }



\begin{abstract}
Here we present an in-depth  analysis of  the periodicity of  the continua and broad emission lines of  a quasar  E1821+643.
We applied non-parametric composite models, the linear sum of stationary and non-stationary Gaussian processes, and quantified contribution of  their periodic parts.
We found  important qualitative differences among the three  periodic signals.
Periods of    $\sim 2200$ and $\sim 4500$ days appear in both  continua 5100 \AA \, and 4200 \AA \,,  as well as in the  H$\gamma$ emission line. 
Their
integer ratio  is nearly  harmonic  $\sim \frac{1}{2}$, suggesting the same physical origin.
We discuss the  nature of  these periods, proposing that  the system of   two objects in dynamically interaction can be   origin of  two largest periods.  
\end{abstract}

\keywords{(Galaxies:) quasars: supermassive black holes; Methods: data analysis}


\section{Introduction}

Investigation of active galactic nuclei (AGN) variability overpowers any other observational technique
in giving an  insight into   their  structure even  on the smallest scales.
Among  the most interesting related issues  is the quest for periodicity, because it could be a signature
of  supermassive black hole binary \citep[SMBHB,][]{LR02}.
Periodicity of AGN  has been in the spotlight  since the work of \cite{Si88}, where was noticed a regularity in the historical optical light curve of the blazar OJ 287.
 The  large flare events occurred 
12 years in the light curve of the blazar, during the period of almost one century of observations.


Up to a few years ago  just some AGNs 
 were considered as  long-term (or quasi\footnote {Some models of multiple black hole systems (e.g. \cite{Sund97}) have shown that perturbations in such systems may not have strictly periodic character, so
we can think of them as quasi (pseudo) periodic.
}) periodic objects
in optical and UV continuum \citep{HT02, F02, La99, Val08, G15a}.
\cite{G15b} made systematic search for the long-term variations of continuum in large sample of quasars and found 
more than one hundred candidates. Thorough analysis of  these results  \citep[see][]{V16}   warned community  about critical interpretation of derived periodicities.
Namely,  AGN periodical flux   variability   
may be explained by   multiplex and violent  
physical mechanisms. We list, as an examples, several possibilities  \citep[for the first three see][ and references therein]{A15}:  pulsational accretion flow instabilities; jet precession, jet rotation,
  helical   jet structure;  an accretion-outflow coupling mechanism; orbiting disk hot spots \citep{Gui83}.
   Also, the last two mentioned  processes can be presented in SMBHB systems too, distorting and hiding periodic signals.
Accordingly, it becomes very important   to be able to  reveal the periodic  signals in AGN  time series  from the noise that is   hiding them. 

Up to now investigations of AGN led to  only few non-controversial explanations of  SMBHB \citep[e. g.][]{PL12, Bon12}. 
Such pseudo-periodic  signals   can be used
in order to predict the epochs of future flux flares within known confidence  limits. This  would  allow us
 to  confirm or discard  multiple black-hole candidates  and plan future
observations based on the expected activity phase of the source.


 Among the  binary black hole candidates is a quasar   E1821+643. \cite{Sh16}, hereafter Paper I,  presented  the results of  the first long-term (1990-2014) optical spectroscopic monitoring of this object.
In their study, application of the standard Lomb - Scargle method \citep[LS,][]{L76} suggested a possible periodicity in the continuum and broad emission line light curves, which may be caused by orbital motion. 
Here  we extended their periodicity analysis  and  investigate the signal/red noise distinction in the continua and emission lines time series of this object.
We applied composite non-parametric models, i.e. the linear combination of  simple Gaussian processes, 
to the  continua at 5100\, \AA,  4200\,\AA,  H$\beta$ and  H$\gamma$   emission lines. The aim of this paper is twofold: first, to investigate in depth periodic signals 
in the light curves of this object  and second, to provide more clues on its broad line region (BLR) geometry and  its  possible SMBHB  nature.

 A short description of used observations and  insight into Gaussian processes are presented in the section  Data sets  and Methods.
In the section Results and Discussion we display  Gaussian processes models of our light curves, their inferred periods and discuss their implications on 
rough geometrical characteristics of BLR of  E1821+643.

\section{Data sets and Methods}

 We examine four diverse time series containing  the integral fluxes of the broad  H$\beta$, H$\gamma$ emission lines and  the fluxes for the continuum 
at the rest frame wavelength of $5100$\, \AA\, and $4220$\, \AA. As the temporal coverage and sampling of  data sets  are important 
for an extensive search for a periodic behavior of  an AGN, reader can find this analysis  in   Paper I.


AGN time series are usually  sparse and irregularly sampled. Even if  each time series is defined over a common continuous time interval, they
can be defined on different collection of time points due to irregularity. Also, the distribution of intervals 
between time points are  usually not uniform and the number of observations in different time series can be different.
Learning any characteristic in such setting is difficult.


It has been shown that the Gaussian process \citep[GP][]{RW06} naturally  conforms to sparse and irregular sampled time series.
GP are suitable to 
 AGN light curves, since in general they are  complex and with no simple parametric form available. In such situation our  prior knowledge about  right model  can be limited  to some   general facts.
 For example, we may just know   
 that  our observations originate from  underlying process which is discrete, non-smooth,  that has variations over certain characteristic time scales and/or  has typical amplitudes.
Surprisingly, it arises naturally to work  with the infinite space of all functions that have  such general characteristics.
As these functions are not characterized with explicit sets of parameters, these techniques are called non-parametric modeling.
Since the dominant tools for working with such model is probability theory, they are recognized as Bayesian non-parametric models, and particular member
 of such models are   GPs. 
  Moreover, AGN variability  amplitude is large and their luminosity is always positive,  allowing us to  perform calculations using variable  $\mathrm{l(t)}=\ln \mathrm{(Flux(t))}$,
  which  can be assumed to be GP \citep{NB15}. 

 The simplified version of GPs (namely continuous time first-order autoregressive process (CAR(1)))  has been used to model AGN continuum light curves \citep[see][and reference therein]{Pan14}.
These works used  Ornstein - Uhlenbeck (OU) covariance function:
  \begin{equation*}
K_{\text{OU}}(\Delta t)=\frac{\hat{\sigma} ^{2}}{2\alpha_{0}}\exp^ {-\alpha_{0}\Delta t}
\end{equation*}
where $\frac{1}{\alpha_{0}}$ is characteristic time scale,  $\hat{\sigma}^{2}$ is variance of driving noise, and $\Delta t$ is the time separating two observations
\citep[see][]{Ke14}.
(Note here that this kernel can be written also in the form
 \begin{equation*}
 K_{\text{OU}}(x,x')=\sigma^{2}\exp {\Big(}-{\frac {|x-x'|}{l}}{\Big)}, 
  \end{equation*}
 where the parameter $\sigma^{2}$ is the variance and $l$ is the characteristic length  scale of the process, i. e. how close two
 points $x$ and $x'$ have to be  to influence each other. 
  The OU process is a stationary GP.
  Since the information about the underlying nature of AGN light curves is crucial
for probing their variability, \cite{Ke11}, \cite{An13} and \cite{Zu13} 
modeled the light curves as a parameterized stochastic process  with 
composite covariance functions. Following the same direction of investigations, we applied  some of stationary and non-stationary composite covariance functions.  As they are  a linear combination of other
 simpler covariance functions, we are able to  incorporate insight about  periodicity, red noise  and  nonstationarity of underlaying processes.

   In our research we employed  two kind of kernel families: stationary -
  already mentioned OU precess, the squared exponential (SE):
     
     \begin{equation*}
    K_{\text{SE}}(\Delta t)=\frac{\hat{\sigma} ^{2}}{2\alpha_{0}}\exp^ 
{-(\alpha_{0}\Delta t)^{2}},
    \end{equation*}
  \noindent  which alternative representation is
     \begin{equation*}
        K_{\text{SE}}(x,x')=\sigma^{2}\exp {\Big(}-{\frac 
{||x-x'||^{2}}{2l^{2}}}{\Big) },
         \end{equation*}
\noindent rational quadratic (RQ)
\begin{equation*}
  K_{\text{RQ}}(\Delta t)=\frac{\hat{\sigma} 
^{2}}{2\alpha_{0}}(1+\frac{\alpha_{0}^{2}|\Delta
  t|^{2}}{\alpha})^{-\alpha },  \text{\,with\,}  \alpha = 2;
   \end{equation*}
\noindent with an alternative form
\begin{equation*}
  K_{\text{RQ}}(x,x')=\sigma^{2}(1+\frac{|x-x'|^{2}}{\alpha l^2})^{-\alpha 
}  \text{\,with\,}  \alpha = 2;
   \end{equation*}
  \noindent
non-stationary -- the standard periodic kernel (StdPer)
     \begin{equation*}
   K_{\text{StdPer}}(\Delta t)=\frac{\hat{\sigma} ^{2}}{2\alpha_{0}}\exp 
\Big(-{\frac {2\sin ^{2}({\frac {\pi(\Delta 
t)}{P}})}{(\frac{1}{\alpha_0})^{2}}}{\Big)}
       \end{equation*}
\noindent  with alternative version
     \begin{equation*}
   K_{\text{StdPer}}(x,x')=\sigma^{2}\exp {\Big(}-{\frac {2\sin ^{2}({\frac 
{\pi(x-x')}{P}})}{l^{2}}}{\Big)};
       \end{equation*}
\noindent        and Brownian motion (Brw, red noise or Wiener process)
           \begin{equation*}
           K_{\text{Brw}}(\Delta t)=\frac{\hat{\sigma} 
^{2}}{2\alpha_{0}}\text{min}(\Delta t),
              \end{equation*}
\noindent which alternative form is

         \begin{equation*}
           K_{\text{Brw}}(x,x')=\sigma^{2}\text{min}(x,x').
              \end{equation*}

The SE is infinitely differentiable allowing GP with this covariance function to generate functions with no sharp discontinuities.
 An alternative to the  sum of SE kernels is RQ. It was shown that this kernel is equivalent to  summing  many SE kernels with different scales. This
produces variations with a range of time scales. Parameter $\alpha$ is roughness, it determines the relative weighting of large and small scale variations. Under condition  $ \alpha\to\infty$, 
the RQ becomes identical to SE. In StdPer kernel hyper parameters $P, l$ correspond to the period and length scale
of the periodic component of the variations. In this case $l$ is related to $P$ so it has no dimension.

We applied  Brownian motion because it  is, in a certain sense, a limit of rescaled simple random walks.  
Many stochastic processes behave, at least for long ranges of
time, like random walks with small but frequent jumps. The argument above implies
that such processes will look, at least approximately, on the proper  time scale,
like Brownian motion.
It  is very  important when probing the periodicity of
light curves, to take into account the red noise. It  is often presented in the AGN continuum and broad emission lines. Such noise can produce spurious signals in a power spectrum of periodogram. We point out that 'red noise' variability of AGNs originates  in 
physical processes of the source and it is not result of measurement uncertainties.
 The red noise simulations were produced by OU process having  the variance of real light curves.

Since all of these base kernels are positive semidefinite kernels (i.e. the valid covariance functions can be defined) and 
 closed under addition and multiplication, we can create richly structured and interpretable kernels with their  summation and/or multiplication.
By summing kernels, data can be modeled  as a superposition of independent functions, representing different structures.
In the case of our time series models, sums of kernels can express superposition of different processes operating at different scales.
Since we are applying these operations to the covariance functions, the composition of even a few base kernels can catch complex relationships
in data which are not of simple parametric form.
 This enable us to  search for the structure in our data, i.e. periodicity/red noise,  and capture them.
   We determine the number of  probing  and valid periods  by utilizing so called 'a two tier approach' of \cite{Vl05}.
On the first tier the period candidates are extracted from the periodogram analysis. On the second tier, these periods are verified by other method (these authors used 
autocorrelation function). In our case, on the first tier we extracted three candidate periods (see Paper I) by means of periodogram analysis.
Here, on the second tier, we put these three candidate
periods  into composite GP models. If these models
provide better modeling result than their non periodic parts  and if the optimized
values of periods are close to the candidate values (within a few hundred of days) then
we can consider these values as a valid periods.

For our analysis,  we  used the  following kernels:
 
 \begin{align} 
\mathrm{ SE}+\sum_{ \mathrm{i}=1,3} \mathrm{StdPer}_\mathrm{i}\\
\mathrm{RQ}+\sum_{ \mathrm{i}=1,3} \mathrm{StdPer}_\mathrm{i}\\
\mathrm{Brw}+\sum_{ \mathrm{i}=1,3} \mathrm{StdPer}_\mathrm{i}
 \end{align}

\noindent where $\mathrm{StdPer}_{ \mathrm{i}},  \mathrm{i}=1,2,3$ are periodic kernels  for searching for specific  periods Per$_\mathrm{i}, \mathrm{i}=1,2,3$.

The structure of  above defined kernels' covariances can be visualized  by  'heatmap'  of  the
values in the covariance matrix. 
For  demonstration purpose only, arbitrary  one dimensional  input space has been discretized equally, and the covariance matrices of Eq. 1, 2 and 3 are constructed
from this ordered list (see  Figure 1). The region of  high SE and RQ covariance   is depicted as
 a diagonal  white  band, reflecting the local stationary nature of these two kernels. Increasing the lengthscale $l$  increases the width of the diagonal band,
since points at larger distance from each other become correlated.

 \begin{figure}[t]
\includegraphics[width=0.48\textwidth]{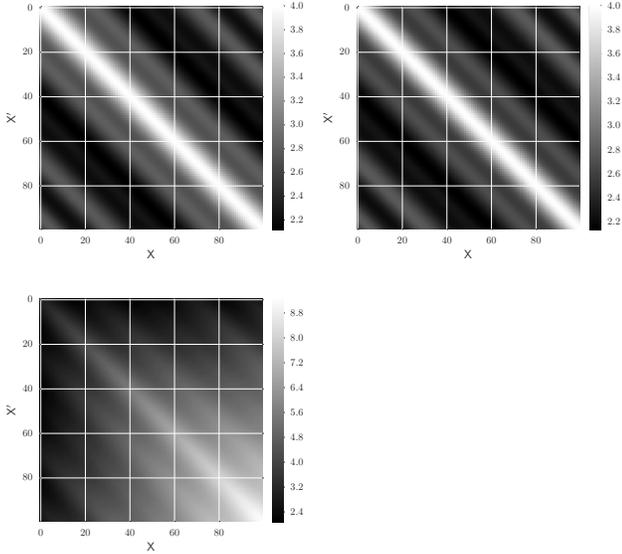}
\caption{%
The prior covariance matrices associated  with  composite  kernels  given by equations 1 (top left), 2  (top right), and 3 (bottom) are calculated from  100 artificial equidistant  points. 
The lightest shade  indicates the highest  and the darkest shade  lowest  covariance.
} 
\end{figure}

Since we summed  periodic and non-periodic kernels it is important to quantify the periodicity of  inferred composite models.
 For this purpose, we applied the method of  \cite{D16}  for the first time 
in astronomy. According to their prescription, let   $y_p$ and $y_a$ be the periodic and non-periodic 
components of our  model calculated on random variable of  input data points. To quantify the periodicity signal, we calculate periodicity ratio  as:

\begin{equation}
\mathrm{S}=\mathrm{\frac{Var(y_p)}{Var(y_{p}+y_{a})}}
 \end{equation}
Furthermore,  $\mathrm{S}$  cannot be viewed as a percentage of the  periodicity of the signal rigorously since  $\mathrm{Var(y_{p}+y_{a})} \neq  \mathrm{Var(y_{p})+ Var(y_{a})}$.
Consequently, $\mathrm{S}$ can have values greater than 1.

For selecting a model from a set of candidate models, which is best supported by our data, the Bayes factor (BF) \citep{J61,KR95} is used.
The BF of two models $ \mathrm{Model}_{i}, \mathrm{Model}_{b}$   is just ratio of their marginal likelihoods ML($\mathrm{Model}_{i}$),  ML($\mathrm{Model}_{b}$) derived using the same dataset, and it includes all information about a particular Bayesian model. BF is preferred
for model selection criteria because they contain the information about a model prior.
For interpreting the values of BFs we used half-units of the logarithmic metric  ($\mathrm{LBF}=\log_{10} (\mathrm{BF})=\log_{10} \mathrm{ML}(\mathrm{Model}_{i})-\log_{10} \mathrm{ML}(\mathrm{Model}_{b})$), since 
we have already  calculated the marginal log likelihood of our models.  
The interpretation of LBF  is that when $0<\mathrm{LBF}<0.5$, there is an evidence against  $\mathrm{Model_{i}}$ when compared with $\mathrm{Model_{b}}$, but it is only worth a bare mention.
When $0.5\leq \mathrm{LBF}<1$ the evidence against $\mathrm{Model_{i}}$ is definite but not strong.
For $1\leq \mathrm{LBF} \leq2$ the evidence is strong and for $\mathrm{LBF}>2$ it is decisive. In our case we will compare models  to the base model - red noise  $\mathrm{Model_{b}=M_{OU}}$ which  is generated from  OU kernel.
Results  in this work  have been calculated with 1.0.7 version  of the Pythonic Gaussian process toolbox GPy (\url{http://github.com/SheffieldML/GPy}).

\section{Results and Discussion}

 As our GP models have a number of  hyperparameters (the covariance functions are composites,  see Table 1), we first assigned a prior values to some of  hyper parameters, obtained
 from our analysis of light curves in Paper I. In assigning  informative  priors to   three  periods in our  GP models, we used  priors  of  4500, 2000 and 1400 days (from the LS periodogram analysis), constraining that the most important periods were on the order of thousand days, rather than on  larger or smaller scales (e. g. seconds or millennia).  So the range of priors to the periods were [0,1400],[0,2000] and [0,4500] days.
 

From the heteroskedastic  OU process and CARMA model \citep{Ke14}  random light curves were  sampled to a 
irregular time intervals of real light curves, in order to
 to test if periodicity found by LS method  is  the product of random variations.
In such a way, we also compare  their effects as a base red noise models.
The reason for trying this two types of models  is a radio-loud, X-ray nature of  E1821+643, due to which
its light curves may not  be well described by OU models. Moreover, \cite{Ko16}
has shown that OU models can be a degenerate descriptors. A CARMA models might serve better
for this purpose, particularly as the red noise terms in them
can produce quasi-periodicities which are missing from simple OU models. For the purpose
of CARMA modeling we used Brandon Kelly carma\_pack Python package available on
\url{https://github.com/brandonckelly/carma_pack}. CARMA models lightcurve as sum of (deterministic) autoregression plus (random) stochastic noise.
The CARMA model order input is optional. We  automatically choose the CARMA order (parameters (p,q)) by minimizing the AIC (Aikake Information Criterion).
Also, the residuals from the one-step-ahead predictions standardized by their standard deviation were uncorrelated. Random light curves were sampled with
built in functions in GPy and farm\_pack packages.

We recalculated periods of each real and simulated curve (see Figure 2) by means of  
Bayesian formalism for the generalized LS periodogram \citep[BGLS][]{M15}, which gives
  the probability  of  signal's existence  in the data. 
It can be seen that   powers  of BGLS peaks of  OU and CARMA 'red noise'   curves 
are asymptotically close to each other as well as to the   continua 4200 \AA \, 5100 \AA\ and the H$\gamma$ line (Figure 2). 
 
Due to  this fact  we will  consider as base red noise  model OU.  Also, in our calculations will be included models  given by Eq. 3 and

\begin{equation*}
  \mathrm{PM_{K}=K+\sum_{i=1,3} StdPer_{i}, K=OU, Brw}.
 \end{equation*} 
  Models  $\mathrm{PM_{OU}, PM_{Brw}}$  consist of a set of  signals ($\mathrm{StdPer_{i}}$) superimposed on the red noise ($\mathrm{Brw,OU}$).
These models allow us to disentangle  signals in the light curves from the 'red noise' background in which they are immersed.

 \begin{figure}[t]
\includegraphics[width=0.48\textwidth]{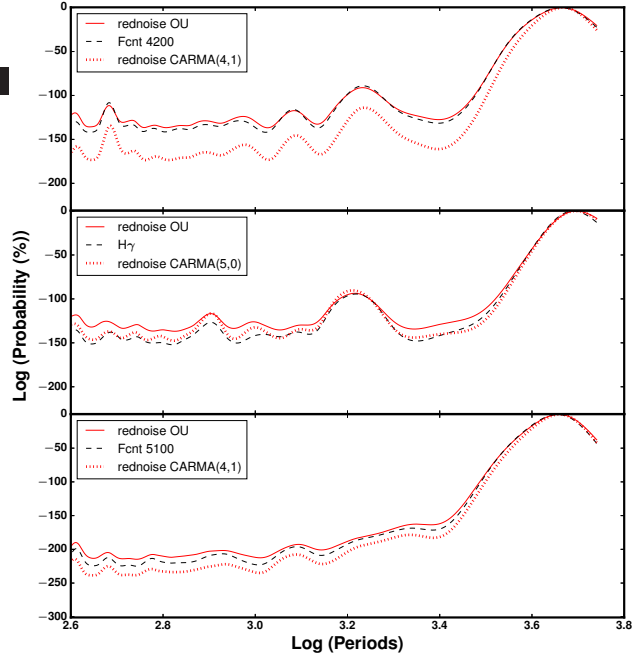}\\

\caption{%
Bayesian periodograms of the simulated 'red noise' and real light curves on a logscale.} 
\end{figure}

 We initially set default hyperparameters of variance to the variance of light curves and 
lengthscale between 0.5 and 100 days.
The variance determines the average distance of our function away from its mean. 
Initially  we expect to have the same variance as our light curves.
 We chose  range of lengthscales between 
the minumum of sampling times of observations ($\sim$ 0.5 days) and twice of mean of sampling times 
($\sim$ 100 days). The lengthscale determines the length of the 'wiggles' in our function, 
so with larger  lengthscale than 100 days  the function would be smoother.

For each light curve, the values of these parameters are estimated using maximum likelihood.
The assigned  values of the models parameters may not be relevant 
for the current data (e. g. the confidence intervals of models can be  too wide). So, we  optimized them  by maximizing the marginal log likelihood of the observations,
 using Broyden-Fletcher-Goldfarb-Shanno (BFGS) algorithm,  which is the option of the GPy tool. It automatically
finds a compromise between model complexity and data fit.

 To address the  heteroskedasticity of data errors we have applied
homoskedastic GP regression on logarithmic transformed  dependent variable  
heteroscedastic GPy regression on non-transformed data.
 Both methods provide almost the same results, which is clearly seen in the case of three periodic composite models (Table 1 and Table 2).
 
The one periodic component models (Table 3) are not favored since S are large while
LBF are small and vice versa for heteroscedastic regression (S$_{het}$ are small and LBF$_{het}$ is large).
Despite  two periodic components models (Table 4) have large S and S$_{het}$
they are not favored since their mean values of LBF and LBF$_{het}$ are smaller 
than corresponding values in Table 1 and Table 2 respectivelly.
So three periodic component models inferred from homoskedastic (Table 1) and heterskedastic regressions (Table 2) are dominant, providing almost the same results. 

For simplification purpose and following prescription of \cite{NB15}, we will refer on the results from Table 1 in further discussion.

In  Table 1, we compare and contrast 
quantitatively how well each model captured the pattern of periodicity and random noise of  each light curve. 
  Before any quantification of  the difference between models, from simple visual inspection of Figure 3  can be seen that  there is  more uncertainty around the predictions of M$_\mathrm{OU}$ then other models.
Actually,  in regions of sparse data, the predictive distribution is unchanged from the prior, whereas in data dense regions, the
uncertanity  is smaller.
 This   suggests worse performance of  GP without periodic components.
 The predictions improve notably as more structure is added to models (i. e. periodic kernels).
 The similar   characteristics,   but not so pronounced,  can be seen in the models of continuum 4200 \AA\, (Figure 4).
 Also,  Table 1 shows that even  from combination of  three periodic kernels the learned model 
 will select two periods. This is not the case when we used M$_1$, M$_2$ models for the  continuum 5100 \AA \, and 
 PM1$_\mathrm{OU}$ model for the continuum 4200 \AA.
 Perhaps explanation lies  in the very nature of these signals (i.e. the shortest periodic signal is weakest).
 
 None of our composite models were able to represent the  H$\beta$ line.  An example of such failure is depicted on Figure 6.  We attribute  it to the  loss  of  periodic signal. Furthermore, in Paper I was shown that the time delay of    H$\beta$  is larger
than the H$\gamma$ emission line, meaning that the   H$\beta$  line originates at larger radii than H$\gamma$. Combining previous facts  (failure of H$\beta $ modeling and its larger distance of
origin),  it seems 
that at  the H$\beta$  line distance some chaotic   processes  can exist (like outflows) destroying all periodic signals. We discuss reasons for this scenario  in the Section 3.1.

Moreover our learned models did not select a single period which is the least common multiple of three (or two larger) periods. 
 This fact shows that such model with one period cannot  reproduce the  light curves while the models with two periods  can produce
them  very well. The results imply that the contributions from the two periods  are significant, and the flare-lake events reported in Paper I are not random but have
the same physical origin.
 
 \begin{figure*}[t]
\includegraphics[width=0.7\textwidth]{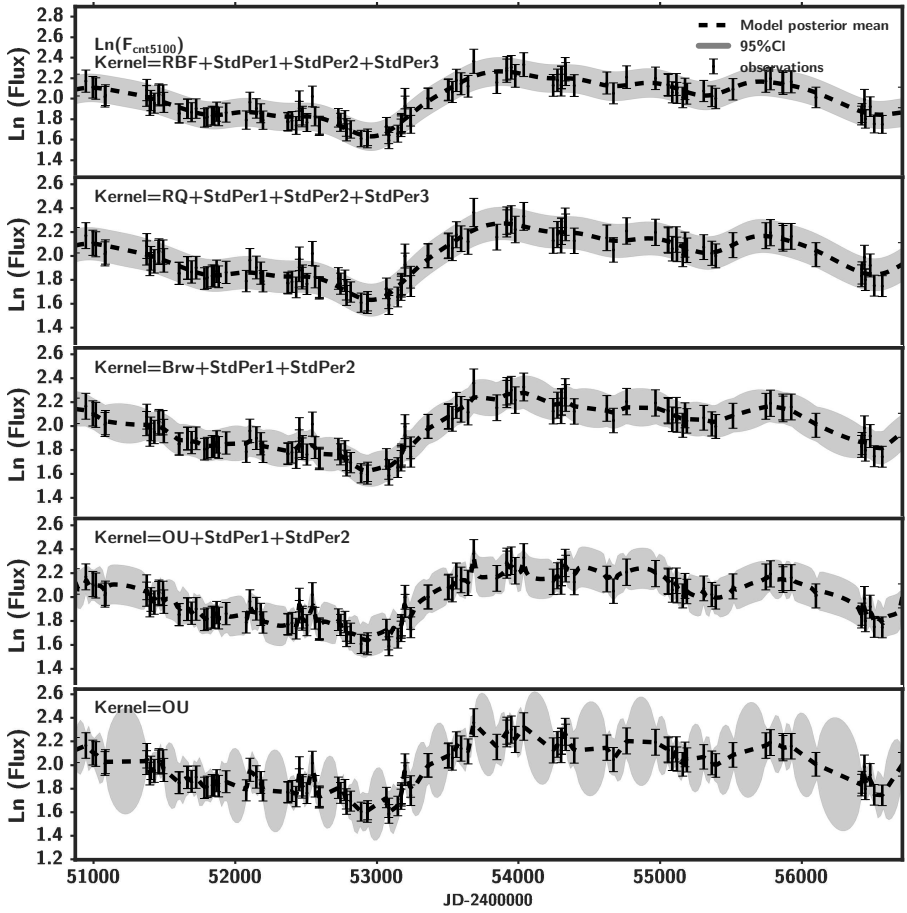}\\

\caption{%
Comparison of GP posteriors (dashed line - posterior mean function, shaded region - $95\%$ confidence interval)  of the continuum at 5100 \AA\, 
(denoted on each plot as markers with error bars) for different kernel composites. We used natural logarithmic values of fluxes. Composite of  Brownian and three periodic kernels provide a richer form of mean function, which explains better data.} 
\end{figure*}

 The contribution of  periodic part  of each model is estimated  with the ratio S given by Eq. 4. For all models the  periodic part is strong ($\mathrm{S}\geq1$).
 We emphasize that  flare like events are associated with GP mean curve, 
 which ensures that the behavior of the GP models cannot simply be interpreted as overfitting due to added periodic kernels.
 Also, Table 1 shows that the  LBF exhibited preference for the models with periodic parts. 
 Factor values larger than 2 on the log$_{10} $ scale gives decisive evidence in favor of  composite models (with periodic parts).
 Thus, all models  (except M$_3$)  expressed decisive evidence in favor of  composite models.
However, M$_3$ model of H$\gamma$ (Figure 5) and the continuum 5100 \AA\, still indicate that the data express substantial to  strong evidence in favor of composite models. 
The mean periods averaged over all models are given in MP row of Table 1.
The  shortest  value of averaged  periods is  comparable with the consecutive intervals between 4 peaks in R band photometric light curve found in Paper I  (see their  Fig.1).
While medium period almost equals the intervals between the first flare and third, and second and last flare.
The largest period is somewhat greater than the interval between the first and the last flare in R band curve since its monitoring period is shorter than optical spectroscopic light curves.
However it  equals the interval between the two local maximum in optical fluxes observed around MJD 51500 and 56000 (see Figure1-3,5 in Paper I).
Association of  $\sim $ 4500 days period with ocurence of  two local maximum in optical light curves as well as 
its comensurable relation with period of $\sim$ 2200 days indicate its physical relevance, despite of 
observational coverage of  5700 days \citep[see][for statistical conditions of reliable period detection]{V16}.

\subsection{The nature of periodicity in E1821+643}

Here we will analyze possible physical origin of inferred periods. It is interesting to see do we have enough information to discern 
whether  the detected periods can  arise from the  orbital motion of   distinct sources of radiation (such as bright spots) within the accretion disk
 or  distinct physical objects in mutual dynamical interaction (which can be also binary black hole system).
 Through variety of  numerical simulations it has been determined that configurations of SMBH binaries assume
circumsecondary ("mini") disk  around secondary component, beside circumbinary disk. Also, there are scenarios 
where mini-disk is seen as the main source of the emission. One of signature
of mini-disk presence would be ripples and/or oscillations in the broad FeK$\alpha$ line as suggested by \cite{McK13}.
However, these signatures will be possible to detect with future observational missions.
The numerical study of \cite{N12} shown that if the accretion rate  to the
inner regions of circumbinary disk were comparable to that of ordinary AGN, the surface density, and therefore the luminosity,
of such a circumbinary disk could approach AGN level. Most strikingly, the luminosity should be
modulated periodically at a frequency determined by the binary orbital frequency and the binary
mass ratio. Since E1821+643 is one of the most luminous quasars and we detected periodical signals in optical
light curves, we will put our discussion within the scenario of luminous circumbinary disk, being aware  of other
possibilites.

 The reason to consider the first  possibility is that broad line profiles of this object have  unusual shapes (thorough line shape  analysis is given in Paper I). 
 If the accretion disk is non-axisymmetric, changes in the line shapes will be caused by the disk features orbiting  on dynamical time scales.
 The dynamical time scale (orbital time) of the light emitting region  at the radius  $r$ of the disk can be approximated by
 
 \begin{equation*}
  t_{dyn}\simeq 5\times 10^{-3}\left ( \frac{M}{10^{8}M_{\odot}} \right )\left ( \frac{r}{r_{g}} \right )^{\frac{2}{3}}
   \end{equation*}  
 \begin{equation*}   
r_{g}=\frac{GM}{c^2}
  \end{equation*}    
\noindent  where  $ r_{g}=\frac{GM}{c^2}$  is the gravitational radius around a black hole of mass $M$, $G$ is the gravitational constant and $c$ the speed of light.
By means of  this relation we can estimate orbital time of the  H$\beta$ and H$\gamma$ broad line  emitting region.
Using  their time lags  (120 ld for H$\beta$ and 60 ld for H$\gamma$, see Table 10 in Paper I), we estimate that corresponding values of  $t_{dyn}$ to their  radii of origin
are  $\sim 11.3$ yr  and $\sim 7.1$ yr, respectively.  These orbital times within the disk are  close to two mean largest periods detected by GP (Table 1). However, 
GP models failed to detect  any period in the H$\beta$ line, which prevents us to explain two largest detected periods as the orbital motion within the disk of features such as bright spots. 
Failure of GP models to extract periods from the H$\beta$ line can be explained by its deterioration.
The degree of  its disfiguration  is easily diagnosed by comparison of    the  mean profiles of H$\beta$ and H$\gamma$  (see Figure 17 in Paper I).
 The mean H$\beta$ profile exhibits more extended red wing then the mean H$\gamma$.
It can be caused by different mechanisms. An additional emission in the far wing of the mean H$\beta$ or even   characteristics of  observational  instruments can result in such degradation of  the H$\beta$ line. 
Also, some processes like outflows can  occur within the  disk. We recall that  \cite{B07} proposed that  a black hole ejection should be 
 uncommon in the aftermath of gas-rich mergers.  In the  gas-rich galactic mergers, torques arising from gas accretion align 
 the spins of supermassive black holes and their orbital axis with a large-scale gas disk. On the other hand, kicking of the black  hole  may happen
  especially in merging galaxies that are relatively gas-poor. So if the kick speed of this object is about  the escape velocity of the host galaxy \citep[as reported by][]{Ro10} then
  we should expect that disk is gas poor and violent processes may contribute to it.

Close similarity between  calculated orbital times  ($t_{dyn}\sim$11.3 yr and 7.1 yr,  corresponding to the radii of H$\beta$ and H$\gamma$ origin within the disk),  and two largest periods detected by GP in both continua (at 4200 \AA\ and 5100 \AA) and the H$\gamma$ line  is not  decisive factor  favoring  bright spots within the disk.
Because of this we consider another possibility that  dynamically related  physical objects (which can be also binary black hole system) can cause periodic signals.

  Two larger periods ($\sim 2200$ and $\sim 4500$ days, see Table 1) can  originate from the real periodic process, 
 because they are   related  commensurably (their  ratio is $\sim \frac{1}{2}$).
 Information about  third  signal of  $\sim 1700 $ days is  weakly presented in both continua  and is   absent from H$\gamma$. 
We note that this period is  incommensurate with respect to the two largest periods. It indicates that
this  signal can be caused by some other process   (like flares) within the accretion disk. 
In such case, 1700 day orbital motion within the disk would correspond to disk's radius  of 2 ld.

In the simplest scenario the two largest periods detected by GP can be attributed to the periods of the orbital motion of  any two   objects in the close mutual dynamical interaction. 

\cite{Ro10} reported that E1821+643 might be an example of gravitational recoil, while resulting SMBH is moving with velocity of $\sim 2100\, \mathrm{km\,s}^{-1}$.
Velocities of such order  of magnitude have been found in many theoretical works. For example, \cite{Gon07} predicted kicks of  $\sim 2500\, \mathrm{km\,s}^{-1}$, \cite{Cam07}  found kicks velocity up to $\sim 4000\, \mathrm{km\,s}^{-1}$ by means of empirical formula, and  \cite{LZ09} obtained  $\sim 3300\, \mathrm{km\,s}^{-1}$ for nearly maximal spins of black holes in the system.
It is believed that  configuration, leading to such events   (with high  velocities kicks),  comprises  of  nearly equal mass circular binaries black holes, with large equal and opposite  spins in the orbital plane.
Morevover, orbital angular momentum and sum of  two black holes'  spin-vectors precess  around total angular momentum during the in spiral (Barusse and Rezzola, 2009).
If the conservation of total angular momentum is valid up to some approximation then the  spin and orbital plane precess at the same frequency.
In this scenario, we can assume that  the precessions of two spin-vectors and orbital angular momentum is  $\sim 12 $ yr , while the orbital period of binary system is  $\sim 6$ yr (see Table 1).
We can use the  following  relation to estimate the rotational period of  black hole \citep[see][]{VV07}:

\begin{equation}
\frac{m_{1}+m_{2}}{m_{2}}=\frac{3}{4}\frac{T_\mathrm{R}T_\mathrm{Prec}\cos\theta}{T^{2}_\mathrm{O}}
 \end{equation}
\noindent where ${m_{1}, m_{2}}$ are masses of  two black holes  in the system, $T_\mathrm{R}$  and $T_\mathrm{Prec}$ are rotational  period of black hole  and precession 
period of its orbit, while $T_\mathrm{O}$ is orbital period. The angle $\theta$ is half-angle of the precession cone.
Assuming  the binaries can be of equal mass ${m_{1}\sim m_{2}}$, and  taking  $m_{1}+m_{2}=2.6\times 10^{9} {M}_{\odot}$ from Paper I, 
$ {T_\mathrm{Prec}}=12.2 $ yr ${T_\mathrm{O}}=5.8 \,$yr then 
we obtain    ${T_\mathrm{R}}\sim \frac{7.4}{\cos\theta}$ yr.  In close binary system $\cos\theta \sim 1$ since  $\theta$ converge to small values.
It is clear that  the period of $4.8 $ yr (see MP row of Table 1) is not related to the rotational period. 
This is one more reason to believe that  it can originate from the random processes within the disk.

 \begin{figure*}[t]
\includegraphics[width=0.7\textwidth]{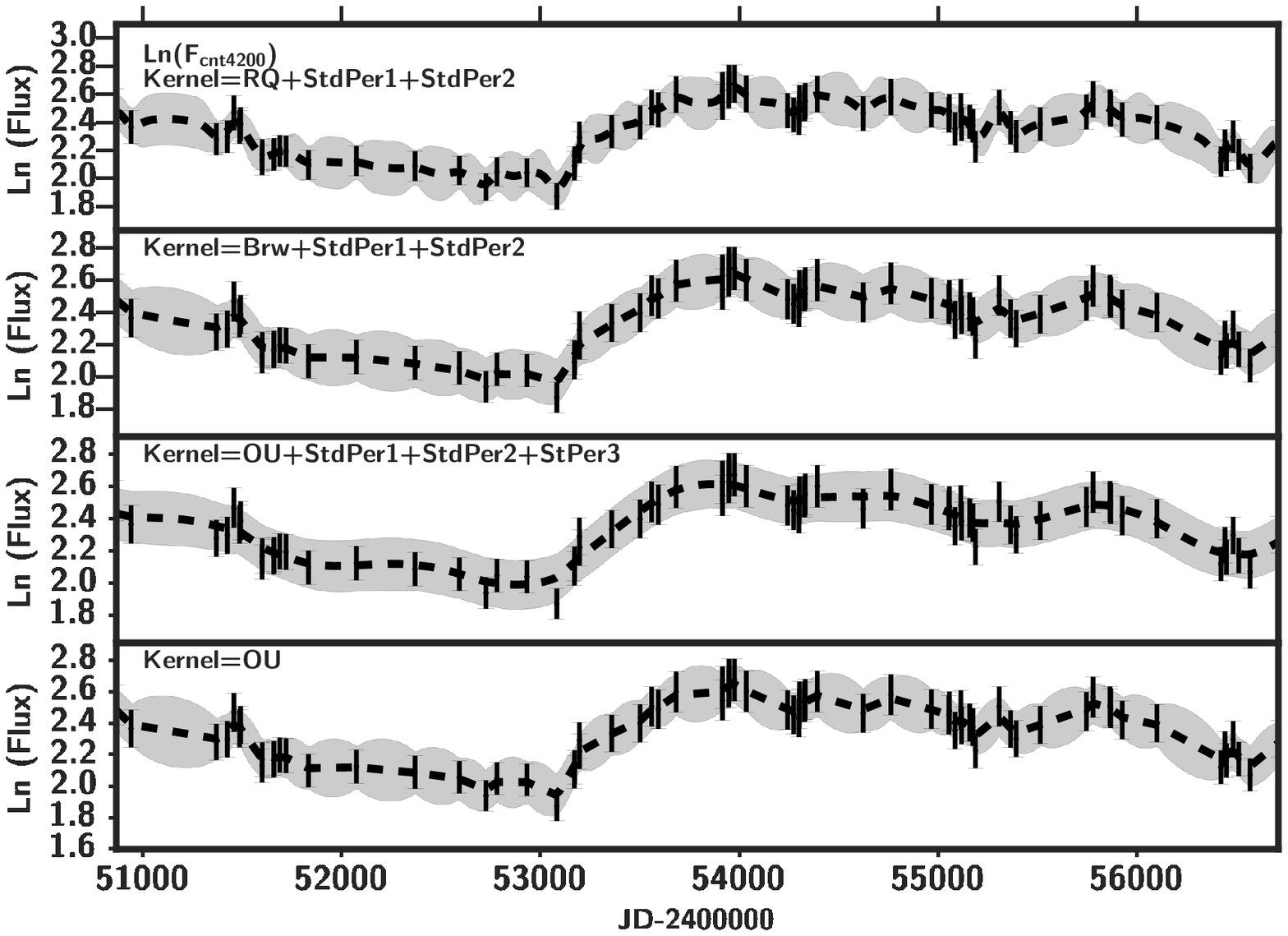}\\
\caption{%
The same as in Figure 3 but for the continuum at 4200 \AA. Note that modeling was possible only with two periodic kernels.
} 
\end{figure*}

Using the third  Kepler's law, we can calculate the radius binary black hole  orbit   $a$:
\begin{equation}
\Big(\frac{T_\mathrm{O}}{1+z}\Big)^{2}=\frac{4\pi^{2}a^{3}}{G(M+m)}
 \end{equation}
Regaining the value  $T_\mathrm{O}=5.8$ yr,  we derived   $a\sim 6\times 10^{14}$ m.
The upper bound of natural gravitational wave frequency is given by  $f_\mathrm{GW}< \frac{1}{4\sqrt{2}\pi}\frac{c^3}{GM}$, where $M$ is the mass of the system
\citep{H03}. In  our case 
$f_\mathrm{GW}\sim4.4 \times 10^{-6}$ Hz. There is no possibility  to measure gravitational waves of such  low frequency using current ground-based instruments.

 \begin{figure*}[t]
\includegraphics[width=0.7\textwidth]{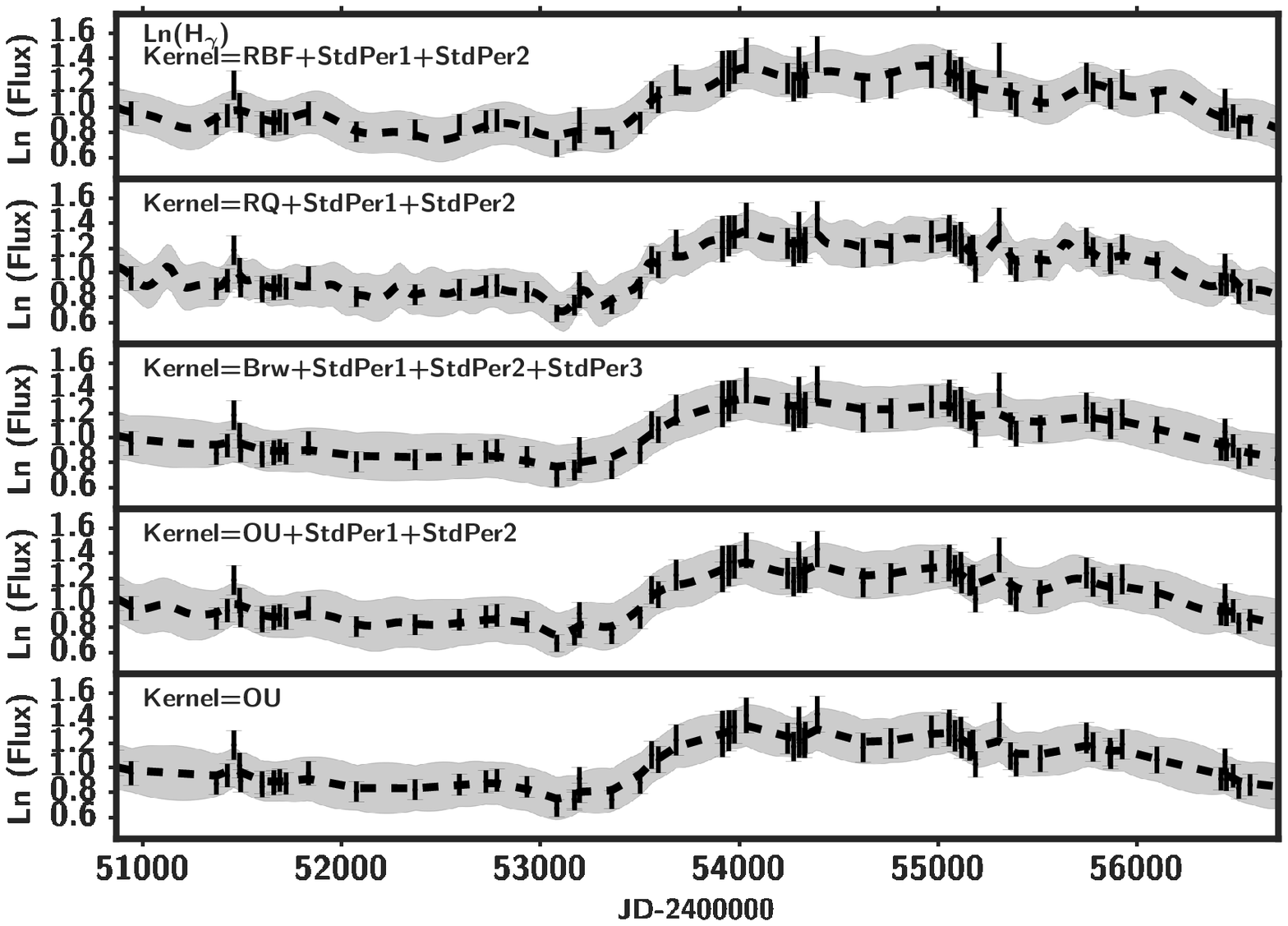}\\

\caption{%
The same as in Figure 3 but for the H$_{\gamma}$ emission line. Note that  stationary kernels composites  includes only  two periodic kernels.
} 
\end{figure*}

 \begin{figure}[t]
\includegraphics[width=0.48\textwidth]{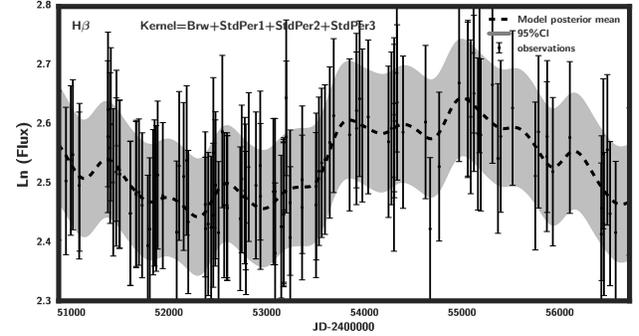}\\

\caption{%
The same as in Figure 3 but for the H$_{\beta}$ emission line and just one composite kernel. The shaded area is enlarged around all  points, proving
that the GP fails to describe daataset. It seems that chaotic processes destroy periodic signals in the H$\beta$ line, preventing
GP models to capture any such pattern in the data.
} 
\end{figure}

\setlength{\tabcolsep}{1pt}
\begin{table}
\small
\caption{Comparison of light curve  models without heteroscedastic errors, using  Bayes factor. In  Lc column are given the light curves, used models are given in column M. Number of parameters used in each model are given in column NP, while inferred periods from models  are given in columns Per$_1$, Per$_2$, and Per$_3$.  LBF column stands for  Bayes factor, ML column stands for marginal log likelihoods of models, S column stands for periodicity ratio of model, Ch column stands for the order of model based on LBF.  In MP row are given averaged periods (in unit of days) from all models, while the brackets denote averaged periods  in the units of  years.   \label{tbl-2}}
\begin{tabular}{@{}p{0.39cm}ccccccccc@{} }
\tableline
Lc & M &NP & Per$_1$&Per$_2$&Per$_3$&ML & LBF& S & Ch\\
      &             &       & $[day]$ &$[day]$   & $[day]$  & & & & \\
\tableline
\multirow{5}{*}{\rot{90}{Cnt$_{5100}$} }   & M$_\mathrm{OU}$\tablenotemark{a}& 3 &- &-&-&-125.76&-&-&-\\

              &   PM$_\mathrm{OU}$\tablenotemark{b}    & 9  &4829 &2345&-&-133.56&7.8&1.23&1\\

         &   M$_1$\tablenotemark{c}    &  12 &4857 &2344&1628&-131.62&5.86&1.00&2\\

                        &M$_2$\tablenotemark{d}&13   & 3498  &2363&1749&-131.04&5.28&1.02&3\\

                        &M$_3$\tablenotemark{e}&  11& 3787& 1758&-&-126.87& 1.1&1.02&4\\
\hline            
\multirow{4}{*}{\rot{90}{Cnt$_{4200}$} } & M$_\mathrm{OU}$& 3&- &-&-&-63.41&-&-&-\\

              &   PM1$_\mathrm{OU}$\tablenotemark{f}& 12  &3852 &1163&1860&-66.36&2.95&1.0&2\\

                        & M$_{22}$\tablenotemark{g}&9  & 4894&2372&-&-69.73&6.32&1.85&1\\

                        &M$_3$&8  &4028&1860&-&-63.9&0.49&1.44&Weak\\
\hline  
\multirow{5}{*}{\rot{90}{H$_{\gamma}$} }& M$_\mathrm{OU}$&3&-  & -&-&-56.75&-\\

             &   PM$_\mathrm{OU}$ & 9  &4717 &2103&-&-58.92&2.2&1.2&3\\

                          & M$_{11}$\tablenotemark{h}& 9 &4978&2159&-&-62.06&5.31&1.00&1\\

                        &M$_{22}$&10  & 5099&2093&-&-61.89&5.14&1.1&2\\
%
                        
                        &M$_3$&11  &4420 &1907&-&-57.69&0.93&1.13&Weak\\
\hline            
          
          \hline  
\multirow{2}{*}{\rot{90}{MP} } & &   &4451   & 2130 \tablenotemark{i}&1746&&&&\\
& &  & (12.2)  &(5.8)&(4.8)&&&&\\
            
\tableline
\end{tabular}
\tablenotetext{a}{OU model of LC are considered as based model}
\tablenotetext{b}{PM$_\mathrm{OU}$ assumes $\mathrm{OU}+\sum_{ \mathrm{i}=1,2}\mathrm{ StdPer}_{ \mathrm{i}}$}
\tablenotetext{c}{M$_{1}$ assumes Eq. 1  }
\tablenotetext{d}{M$_{2}$ assumes Eq. 2}
\tablenotetext{e}{M$_{3}$ assumes Eq. 3}
\tablenotetext{f}{PM1$_\mathrm{OU}$ assumes $\mathrm{OU}+\sum_{ \mathrm{i}=1,3}\mathrm{ StdPer}_{ \mathrm{i}}$}

\tablenotetext{g}{M$_{22}$ assumes $\mathrm{RQ}+\sum_{ \mathrm{i}=1,2} \mathrm{StdPer}_{ \mathrm{i}}$}
\tablenotetext{h}{M$_{11}$ assumes $\mathrm{SE}+\sum_{ \mathrm{i}=1,2} \mathrm{StdPer}_{ \mathrm{i}}$}
\tablenotetext{i}{Period  of 1163 days is not taken into account, \\ since it is beyond two  sample standard deviation below the mean of  Per$_2$.   }

\end{table}





\setlength{\tabcolsep}{1pt}
\begin{table}
\small
\caption{Comparison of light curve  models with three periodic components using  Bayes factor.
Subscript $_{het}$ stands for modeling with heteroscedastic errors. \label{tbl-3}}
\begin{tabular}{@{}p{0.39cm}cccccccc@{} }
\tableline
Lc & M &NP & Per$_{1het}$&Per$_{2het}$&Per$_{3het}$&ML$_{het}$ & LBF$_{het}$& S$_{het}$ \\
      &             &       & $[day]$ &$[day]$   & $[day]$  & & &  \\
\tableline
\multirow{5}{*}{\rot{90}{Cnt$_{5100}$} }   & M$_\mathrm{OU}$\tablenotemark{a}& 2 &- &-&-&119.62&-&-\\

              &   PM$_\mathrm{OU}$\tablenotemark{b}    & 8  &3565 &2810&1151&110.8&-8.22&0.71\\

  
         &   M$_1$\tablenotemark{c}    &  11 &4435 &-&-&274.896&155.28&3.3\\

                        &M$_2$\tablenotemark{d}&12   & 4472 &1731&1989&112.72&-6.89&0.7\\

                        &M$_3$\tablenotemark{e}&  10& 4827& 4721&2252&110.59& -9.02&0.99\\
\hline            
\multirow{4}{*}{\rot{90}{Cnt$_{4200}$} } & M$_\mathrm{OU}$& 2&- &-&-&105.76&-&-\\
              &   PM$_\mathrm{OU}$\tablenotemark{b}    & 8  &3828 &1833&1151&-101.09&-4.763&0.53\\

         &   M$_1$\tablenotemark{c}    &  11 &3847 &1847&1149&100.76&-5.03&1.48\\

                        &M$_2$\tablenotemark{d}&12   & 4910 &1158&-&100.9&-4.86&1.2\\

                        &M$_3$\tablenotemark{e}&  10& 3796& 1154&-&104.2& -1.56&0.36\\

          \hline  
\multirow{5}{*}{\rot{90}{H$_{\gamma}$} }& M$_\mathrm{OU}$&2&-  & -&-& 31.37\\
              &   PM$_\mathrm{OU}$\tablenotemark{b}    & 8  &4978 &2167&-&26.04&-5.33&0.99\\

 
         &   M$_1$\tablenotemark{c}    &  11 &4997 &2167&-&26.04&-5.33&1.0\\

                        &M$_2$\tablenotemark{d}&12   & 4903 &1962&1154&25.11&-6.26&0.9\\

                        &M$_3$\tablenotemark{e}&  10& 5006& 2202&1551&25.9& -6.3&1\\
\hline            
          
          \hline  
\multirow{2}{*}{\rot{90}{MP} } & &   &4466 \tablenotemark{i}   & 2159&1485&&&\\
& &  & (12.2)  &(5.92)&(4.07)&&&\\

    \hline            
\tableline
\end{tabular}
\tablenotetext{a}{OU model of LC are considered as based model}
\tablenotetext{b}{PM$_\mathrm{OU}$ assumes $\mathrm{OU}+\sum_{ \mathrm{i}=1,3}\mathrm{ StdPer}_{ \mathrm{i}}$}
\tablenotetext{c}{M$_{1}$ assumes Eq. 1  }
\tablenotetext{d}{M$_{2}$ assumes Eq. 2}
\tablenotetext{e}{M$_{3}$ assumes Eq. 3}
\tablenotetext{i}{Period  of  4435 days is not taken into account, \\ due to  large ML and LBF values.   }
\end{table}

\begin{table*}
\small
\caption{Comparison of light curve  models with one periodic component, using  Bayes factor. 
The names of columns have the same meaning as in Table 1. Subscript $_{het}$ stands for modeling with heteroscedastic errors.  \label{tbl-4}}
\begin{tabular}{@{}p{0.39cm}ccccccccccc@{} }
\tableline
Lc & M &NP & Per&ML&S&LBF & NP$_{het}$ &Per$_{het}$& ML$_{het}$& S$_{het}$&LBF$_{het}$\\
      &             &       & $[day]$ &  &  & & & & && \\
\tableline
\multirow{5}{*}{\rot{90}{Cnt$_{5100}$} }   & M$_\mathrm{OU}$\tablenotemark{a}& 3 &- &-125.76&-&-& 2&-&119.62&-&-\\

              &   PM$_\mathrm{OU}$\tablenotemark{b}    & 6  & 4734&-127.79&1.0&2.02&5& 1693&117.95&0.29&-1.66\\

         &   M$_1$\tablenotemark{c}     & 6  & 2348&-112.289&0.84&-13.48&5& 623&110.77&0.68&-8.9\\

                        &M$_2$\tablenotemark{d}   & 7  & 4529&-126.60&0.63&0.83&6&  4500&113&0.63&-6.62\\

                        &M$_3$\tablenotemark{e}&   5   &1160&-126.39&0.62&6.8&4&  3519&113.03&0.6&-6.58\\
\hline            
\multirow{4}{*}{\rot{90}{Cnt$_{4200}$} } & M$_\mathrm{OU}$& 3&- &-63.41&-&-&2&-&105.76&-&-\\

       &   PM$_\mathrm{OU}$\tablenotemark{b}     & 6  & 1841&-64.36&1.18&0.95&5& 3532&102.63&0.46&-3.13\\

                        &M$_1$\tablenotemark{d}  & 7   &2321&-61.21&1.0&-2.2&6&  2419&105.5&0.35&-0.26\\

                        &M$_2$\tablenotemark{e}&   5   &4562&-64.04&0.95&0.63&4& 4763&102.04&0.9&-3.36\\
           
           &M$_3$\tablenotemark{e}&  5  & 4171&-63.79&1.13&0.38&4&  4763&102.4&0.62&-3.36\\

\hline  
\multirow{5}{*}{\rot{90}{H$_{\gamma}$} }& M$_\mathrm{OU}$&3&-  & -56.75&-&-&2&-&31.37  &-&-\\

       &   PM$_\mathrm{OU}$\tablenotemark{b}     & 6  & 2010&-58.22&1.03&1.47&5& -&-&-&-\\

                        &M$_1$\tablenotemark{d} & 7   &1265&-49.08&4.23&-7.67&6&  2196&30.71&0.38&-1.0\\

                        &M$_2$\tablenotemark{e} & 5   &5062&-58.5&1.02&1.75&4& 5024&27.62&0.48&-4.09\\
           
           &M$_3$\tablenotemark{e}&   5   &4595&-57.38&1.04&0.63&4&  4371&29.34&0.54&-2.37\\

\hline

\tableline
\end{tabular}
\tablenotetext{a}{OU model of LC are considered as based model}
\tablenotetext{b}{PM$_\mathrm{OU}$ assumes $\mathrm{OU}+\mathrm{ StdPer}_{ \mathrm{i}}$, $i=1$}
\tablenotetext{c}{M$_{1}$ assumes $\mathrm{ SE}+ \mathrm{StdPer}_\mathrm{i}$, $i=1$}
\tablenotetext{d}{M$_{2}$ assumes $\mathrm{RQ}+ \mathrm{StdPer}_\mathrm{i}$, $i=1$}
\tablenotetext{e}{M$_{3}$ assumes $\mathrm{Brw}+\mathrm{StdPer}_\mathrm{i}$, $i=1$}

\end{table*}

\begin{table*}
\small
\caption{Comparison of light curve  models with two periodic component, using  Bayes factor. The names of columns have the same meaning as in Table 1. Subscript $_{het}$ stands for modeling with heteroscedastic errors.
   \label{tbl-5}}
\begin{tabular}{cccccccccccccc }

\tableline
Lc & M &NP & Per$_1$& Per$_2$&ML&S&LBF & NP$_{het}$ &Per$_{1het}$&Per$_{2het}$  & ML$_{het}$& S$_{het}$&LBF$_{het}$\\
      &             &       & $[day]$ & $[day]$ &  & & & &$[day]$ &$[day]$&&& \\
\tableline
\multirow{5}{*}{\rot{90}{Cnt$_{5100}$} }   & M$_\mathrm{OU}$\tablenotemark{a}& 3 &-&- &-125.76&-&-&2& -&-&119.62&-&-\\

              &   PM$_\mathrm{OU}$\tablenotemark{b}    & 9  & 3784&1760&-126.86&1.06&1.1&8& 3573&1154&110.76&0.7&-8.85\\

         &   M$_1$\tablenotemark{c}     & 9  & 3505&1789&-130.36&1.07&4.59&8& 2396&3040&306.94&4.96&187.324\\

                        &M$_2$\tablenotemark{d}   & 10  & 4744&1607&-128.11&0.63&1.01&9&2903&1682&113.82&0.96&-5.79\\

                        &M$_3$\tablenotemark{e}&   8   &4415&4699&-126.77&0.99&1&7&  4827&4722&110.59&0.94&-9.03\\
\hline            
\multirow{4}{*}{\rot{90}{Cnt$_{4200}$} } & M$_\mathrm{OU}$& 3&-&- &-63.41&-&-&2&-&-&105.76&-&-\\
             
             &   PM$_\mathrm{OU}$\tablenotemark{b}    & 9  & 3750&1163&-66.09&6.65&2.68&8& 3693&1153&101.41&0.59&-4.35\\

                        &M$_1$\tablenotemark{c}   & 10  & 4890&2373&-63.90&0.99&0.48&9&955&1144&103.23&0.87&-2.53\\

                        &M$_2$\tablenotemark{d}&   8   &4848&2373&-69.72&1.37&6.31&7&  4727&-&103.12&0.95&-2.64\\
 
  &   M$_3$\tablenotemark{e}     & 9  & 4317&2298&-64.44&1.02&1.03&8& 3534&-&102.57&0.53&-3.19\\

     \hline  
\multirow{5}{*}{\rot{90}{H$_{\gamma}$} }& M$_\mathrm{OU}$&3&- &- & -56.75&-&-&2&-&-& 31.37&-&-\\
           &   PM$_\mathrm{OU}$\tablenotemark{b}    & 9  &2648&2081&-60.08&1.03&3.33&8& 4509&1124&26.42&0.89&-4.94\\

                        &M$_1$\tablenotemark{c}   & 10  & 4975&2156&-62.05&0.99&5.30&9&4891&2167&26.04&1.0&-5.33\\

                        &M$_2$\tablenotemark{d}&   8   &4993&2163&-61.91&0.97&5.16&7&  4982&2167&26.04&1.0&-5.33\\
 
  &   M$_3$\tablenotemark{e}     & 9  & 4975&2156&-62.06&0.99&5.31&8& 4982&2167&26.04&0.99&-5.33\\

\hline

\tableline
\end{tabular}
\tablenotetext{a}{OU model of LC are considered as based model}
\tablenotetext{b}{PM$_\mathrm{OU}$ assumes $\mathrm{OU}+\mathrm{ StdPer}_{ \mathrm{i}}$, $i=1,2$}
\tablenotetext{c}{M$_{1}$ assumes $\mathrm{ SE}+ \mathrm{StdPer}_\mathrm{i}$, $i=1,2$}
\tablenotetext{d}{M$_{2}$ assumes $\mathrm{RQ}+ \mathrm{StdPer}_\mathrm{i}$, $i=1,2$}
\tablenotetext{e}{M$_{3}$ assumes $\mathrm{Brw}+\mathrm{StdPer}_\mathrm{i}$, $i=1,2$}

\end{table*}

\section{Conclusions}

We have performed a set of   GP modeling (Table 1) of the continua at 4200, 5100 \AA\,  and  the H${\gamma}$ emission line light curves  of E1821+643.
 Following direction of  mixed covariance stochastic modeling  of   AGN light curves, we used composite GP models consisting  of   Ornstein - Uhlenbeck,  the squared exponential, the rational quadratic, Brownian and non-stationary -- periodic kernels. 
This allows us to quantify periodic signals in the light curves.

We found three strong signals with  periods of  $\sim 12$ yr, $\sim 6$ yr, and $ 4.8$ yr (averaged values over all models).
 The two largest periods are in a harmonic relationship,  which indicates  their  common  physical origin.
 Analyzing  the scenario with orbital motion of bright spots within the accretion disk, we found that $t_{dyn}$ scales corresponding to the radii of origin of the  H$\beta$ and H$\gamma$
 lines are almost equal to the 
 two largest periods detected by GP models. However, the bright spot motion is not favored because  any information about periodic signals is vanished  from the H$\beta$ line.
  Since the H$\beta$ line arises at larger radius than H$\gamma$, it seems that  here   some chaotic processes (like outflow) are presented that are destroying  all periodic signals.

 The third period of  $4.8$ yr  is incommensurate with respect to the  two largest periods, suggesting its different physical origin  (such as flares at the radius  of  2 ld  within the  accretion disk of the system).

So we propose that two largest periods correspond to the mutual dynamical interaction   of   two objects at least.
Also, we analyzed  the subcenario if these  two objects are  two kicked  merging  black holes.
 In such situation, we associated period of  $\sim 12$ yr with the precession 
of  orbital angular momentum of the system and the precession of the   sum of  two black holes' spins during the infall. The orbital period of black hole within this system can be associated with $\sim 6$ yr.
In such setting, the  rotational period of black hole  would be nearly $\frac{7.4}{\cos\theta}$ yr, where $\theta$ is half-angle of  the precession cone.


\acknowledgments
This work was supported by the Ministry of Education and Science of Republic of Serbia through the project Astrophysical Spectroscopy of Extragalactic Objects (176001) and RFBR (grants N12-02-01237a, 12-02-00857a, 12-02-01237a,N15- 02-02101).

\nocite{*}
\bibliographystyle{spr-mp-nameyear-cnd}
\bibliography{biblio-u1}

\begin{thebibliography}{}

 \bibitem[Ackermann et al.(2015)]{A15} Ackermann, M., Ajello, M.,  Albert, A.,  Atwood, W. B. et al., 2015, \apjl, 813, 2, L41
 
  \bibitem[Andrae et al.(2013)]{An13}  Andrae, R., Kim, D. W., Bailer-Jones, C. A. L., 2013, \aa, 554, 137 

 \bibitem[Barausse \& Rezzolla(2009)]{BR09} Barausse, E., \& Rezzolla, L. 2009, \apjl, 704, L40

 \bibitem[Bogdanovi{\'c} et al. (2007)]{B07}Bogdanovi{\'c}, T.,  Reynolds, C. S., \&  Miller, M. C. 2007, \apjl, 661,  L147

\bibitem[Bon  et al.(2012)]{Bon12}Bon, E.,  Jovanovi{\'c}, P., Marziani, P., Shapovalova, A. I. et al., 2012, \apj,  759,  id. 118



\bibitem[Campanelli et al.(2007)]{Cam07}Campanelli, M., Lousto, C. O., Zlochower, Y., \& Merritt, D. 2007, Phys. Rev. Lett., 98. id231102 

\bibitem[Durrande et al.(2016)]{D16}  Durrande, N.,  Hensman, J.,  Rattray, M., \& Lawrence, N. D. 2016,   Peer J Computer Science 2:e50 \url{https://doi.org/10.7717/peerj-cs.50}


\bibitem[Fan et al.(2002)]{F02} Fan, J. H., Lin, R. G., Xie, G. Z., Zhang, L.  et al., 2002, \aap, 381, 1 

\bibitem[Guilbert et al.(1983)]{Gui83}Guilbert, P. W., Fabian, A. C., \& Rees, M. J. 1983, \mnras,205, 593



\bibitem[Gonzalez et al.(2007)]{Gon07}Gonzalez, J. A., Hannam, M. D., Sperhake, U., Br\"ugmann, B., \& Husa, S. 2007,  Phys. Rev. Lett., 98.  id. 231101 

\bibitem[Graham et al.(2015a)]{G15a}Graham, M. J.,  Djorgovski, S. G., Stern, D., Glikman, E.  et al.,  2015a, \nat, 518, 74
\bibitem[Graham et al.(2015b)]{G15b}Graham M. J., Djorgovski, S. G., Stern, D.,Drake, A. J.  et al., 2015b, \mnras, 453, 1562

\bibitem[Hagen-Thorn et al.(2002)]{HT02}Hagen-Thorn, V. A., Larionov, V. M., Jorstad, S. G., \& Larionova, E. G. 2002, \aj, 124, 3031
  \bibitem[Hughes(2003)]{H03} Hughes, S. 2003, Annals of Physics, 303, 142

\bibitem[Jeffreys(1961)]{J61} Jeffreys, H. 1961, Theory of Probability, 3rd ed. Oxford
Classic Texts in the Physical Sciences. Oxford Univ. Press, Oxford. MR1647885  

\bibitem[Kass \& Raftery(1995)]{KR95}Kass, R. E.,  \&  Raftery, A. E. 1995,  Journal of the American Statistical Association,  90, 773

\bibitem[Kelly et al. (2011)]{Ke11}
Kelly, B. C., Sobolewska, M.,  Siemiginowska, A., 2011, \apj, 730,  1,  id. 52, 20 

\bibitem[Kelly et al. (2014)]{Ke14}
Kelly, B. C., Becker, A. C., Sobolewska, M., Siemiginowska, A., Uttley, P., 2014, \apj 788, 1, id. 33, 18

\bibitem[Kozlowski  (2016)]{Ko16} Kozlowski, S,  2016, \mnras  459,  3, 2787


\bibitem[Lainela et al.(1999)]{La99}Lainela, M., Takalo, L. O., Sillanp\H a\H a, A.,  Pursimo, T.  et al., 1999, \apj, 521, 561

\bibitem[Lobanov \& Roland(2002)]{LR02}Lobanov, A. P. \& Roland, J. 2002, 6th European VLBI Network Symposium on New Developments in VLBI Science and Technology, 121

\bibitem[Lomb(1976)]{L76}Lomb, N. R. 1976, \apjs, 39, 447


 \bibitem[Lousto \& Zlochower(2009)]{LZ09} Lousto, C. O., \& Zlochower, Y. 2009, Phys. Rev. D, 79, 064018

 \bibitem[McKernan et al. (2013)]{McK13}
McKernan, B., Ford,K.?E.?S., Kocsis, B., Haiman, Z., 2013, \mnras, 432,1648 


 \bibitem[Mortier et al.(2015)]{M15}  Mortier, A.,  Faria, J. P.,   Correia, C. M.,   Santerne, A., \&  Santos, N. C. 2015, \aap, 573, A101   
 
 
  

\bibitem[Nipoti \& Binney(2015)]{NB15} Nipoti, C., \&  Binney, J.  2015, \mnras, 446, 1820 

 \bibitem[Noble et al.(2012)]{N12}
Noble, S. C., Mundim, B. C., Nakano, H., Krolik, J. H. et al., 2012, \apj, 755, 1,  id. 51, 24 

   
  \bibitem[Pancoast  et al.(2014)]{Pan14} Pancoast, A.,  Brewer, B. J., Treu, T., Park, D.  et al., 2014,\mnras, 445, 3073 
  
  \bibitem[Popovi{\'c} (2012)]{PL12} Popovi{\'c}, L. {\v C}.,  2012, \nar, 56, 74
    
 \bibitem[Rasmussen \& Williams(2006)]{RW06} Rasmussen, C. E., \& Williams, C. 2006, Gaussian Processes for Machine Learning, the MIT Press 

 \bibitem[Robinson et al.(2010)]{Ro10}Robinson, A., Young, S., Axon, D. J., Kharb, P., \& Smith, J. E. 2010, \apjl, 717, 122


\bibitem[Rieger(2004)]{R04}Rieger,  F. M.  2004, \apjl, 615, L5

 \bibitem[Shapovalova et al.(2016)]{Sh16}Shapovalova, A. I., Popovi{\'c}, L. {\v C}.,  Chavushyan, V. H.,  Burenkov, A. N. et al., 2016, \apjs, 222, id. 25

\bibitem[Sillanp\H a\H a et al.(1988)]{Si88}Sillanp\H a\H a, A., Haarala, S., Valtonen M. J., Sundelius, B., \& Byrd, G. G. 1988, \apj, 325, 628


 \bibitem[Sundelius et al.(1997)]{Sund97} Sundelius, B., Wahde, M., Lehto, H. J., \&Valtonen, M. J. 1997, \apj, 484, 180

\bibitem[Valtonen et al.(2008)]{Val08} Valtonen, M. J., Lehto, H. J., Nilsson, K.,  Heidt, J.  et al., 2008, \nat, 452, 851

\bibitem[Vaughan et al.(2016)]{V16}Vaughan, S.,  Uttley, P.,  Markowitz, A. G.,  Huppenkothen, D.  et al., 2016, \mnras, 461,  3145

\bibitem[Vlachos et al.(2005)]{Vl05} Vlachos, M.,  Yu, P,  Castelli, V., 2005,
In:  Kargupta, H., Srivastava, J., Kamath, C., Goodman, A. (eds.)
Proceedings of the 2005 SIAM International Conference on Data Mining, 449

 
   \bibitem[Volvach et al.(2007)]{VV07} Volvach, A. E., Volvach, L. N., Larionov, M. G., Aller, H. D., Aller, M. F. 2007, Astronomy Reports,  51, 450
   
  \bibitem[Zu et al.(2013)]{Zu13}  Zu,ÊY.,  Kochanek,ÊC.ÊS., Koz?owski,ÊS., Udalski,ÊA., 2013, \apj, 765,  2,  id. 106, 7

  
 













 


 
 

   

\end{thebibliography}

\end{document}